\documentclass[11pt,prd,superscriptaddress,onecolumn,nofootinbib]{revtex4-2}
\usepackage{graphicx}
\usepackage{amsfonts,amssymb,amsmath}
\usepackage[a4paper,margin=0.85in]{geometry}
\usepackage{xcolor}
\usepackage{url,hyperref}
\usepackage{braket}
\usepackage{setspace}
\usepackage{tikz}
\hypersetup{colorlinks,linkcolor={blue!55!black},citecolor={red!45!black},urlcolor={blue!45!black},breaklinks=true}
\usepackage{anyfontsize}
\usepackage[caption=false]{subfig}
\newcommand{\be}{\begin{equation}}
\newcommand{\ee}{\end{equation}}
\newcommand{\ba}{\begin{eqnarray}}
\newcommand{\ea}{\end{eqnarray}}

\begin{document}
	 \thispagestyle{empty}
		$\phantom{v}$
	\vspace{2cm}
	\singlespacing
	\noindent{\Large\textbf{Black Holes, Equilibrium, and Cosmology}}
	
	\vspace{2cm}
	Fil Simovic$^\dagger$
	\vspace{0.5cm}
	
	{\small {\it $^\dagger$ 
			School of Mathematical and Physical Sciences, Macquarie University,\\
			\vspace{-1pt}
			\hspace{19pt}Sydney, New South Wales 2109, Australia}}
	\vspace{0.5cm}
	
	{\small Email: \href{fil.simovic@mq.edu.au}{fil.simovic@mq.edu.au}}
	\vspace{1.5cm}

	We trace the origins and development of black hole thermodynamics across the past half-century, emphasizing the framework's relation to classical thermodynamics, and the vital role played by the notions of equilibrium, stationarity, and symmetry. We discuss different interpretations of the first law of black hole mechanics, and assess the validity of its mechanical, process-based interpretation  for evaporating black holes. We bring these ideas to the cosmological realm, and highlight the various difficulties that arise when formulating thermodynamics for black holes in asymptotically de Sitter backgrounds. We discuss a number of proposed solutions and the open questions that arise therein.
	\vspace{3cm}
	
		\begin{center}
	\noindent{\it Awarded an Honorable Mention for the Gravity Research Foundation 
		
		2023 Awards for Essays on Gravitation}
	\end{center}

	\setcounter{page}{0}

	\newpage

	\section*{Introduction}
   \noindent The study of black hole thermodynamics has played a pivotal role in advancing our understanding of quantum gravity, field theory in curved space, and information theory \cite{ashtekar1998a,carlip2001,carlip2014,wald1994a,birrell1982,harlow2016,polchinski2017,strominger2018}. Formulating a thermodynamic description of gravitational objects such as black holes requires a leap in intuition---that geometry itself can possess qualities normally associated with the microscopic constituents of everyday objects. That this assignment is possible challenges historical notions of gravity as an immutable cosmic background, and lends credence to the idea that gravity can be described in terms of some (presumably quantum mechanical) microscopic constituents. Understanding what these microscopic degrees of freedom are is essentially the problem of quantum gravity, representing perhaps the biggest unsolved challenge in theoretical physics today.
   \\
   
   Over the past few decades, research into the thermodynamic properties of black holes has probed the fundamental nature of entropy, led to the development of the holographic principle, the notion of emergent gravity, and more \cite{bekenstein1972,bekenstein1973,bousso2002,bousso2008,verlinde2011a}. Thermodynamics for asymptotically anti-de Sitter (AdS) black holes (with $\Lambda<0$) is an especially rich subject owing to the AdS/CFT correspondence, which posits an equivalence between gravitational theories in AdS spacetimes and conformal field theories (CFTs) on the boundary of those spacetimes \cite{witten1998e,maldacena1999}. The 1--1 mapping of observables and Hilbert spaces between the two theories means that, for example, phase transitions that occur for black holes in the bulk AdS space have counterparts in the boundary CFT theory, allowing for the study of a variety of strongly coupled systems through the black hole phase structure \cite{hawking1983,witten1998a,caldarelli2000b,frassino2014,kubiznak2017}. These studies take place in the arena of {\it equilibrium} thermodynamics; the context that black hole thermodynamics was first formulated in  and sees the widest application today. 
   \\
   
   Equilibrium thermodynamics arises from a long-wavelength, coarse-graining of the underlying statistical physics of a system, where its microscopic degrees of freedom are encoded in a time-independent distribution function $f(p_i,q_i)$ on a phase space $\Gamma$, and macroscopic quantities like temperature $T$, entropy $S$, etc. can be defined through $f(p_i,q_i)$. While the variables of the coarse-grained description do not necessarily have meaningful analogues at the microscopic scale (e.g. the temperature as the velocity dispersion of a large number of particles), they serve to accurately capture the macroscopic properties of the system in some appropriate statistical or thermodynamic limit. A useful description of the system is thus obtained where tracking microscopic degrees of freedom is impossible. The macroscopic quantities satisfy the fundamental relation
   \be\label{1st}
   dU=T\,dS-P\,dV+\sum_i \phi_i\, d\Omega_i\ , \qquad \phi_i\equiv\left(\frac{\partial U}{\partial \Omega_i}\right)_{\Omega_{j\neq i}}\ ,
   \ee
   where $U$ is the internal energy and the $\phi_i$ are various thermodynamic potentials conjugate to the extensive parameters $\Omega_i$, which may be the electric potential $\Phi$, angular velocity $\Omega$, chemical potential $\mu_i$, etc.  This is the first law of thermodynamics, which represents the change in internal energy of a closed system during a {\it quasi-static} transition between two {\it equilibrium} states. Equilibrium is stressed because it is a requirement of the Clausius relation $dQ=TdS$, which guarantees the existence of the state function $S$ \cite{clausius1867mechanical,landau2011}. 
   \\
   
   Unexpectedly, the early 1970s witnessed a number of investigations which revealed that classical black hole solutions of general relativity obey a law analogous to \eqref{1st}. In a series of groundbreaking works \cite{bekenstein1972,bardeen1973,bekenstein1973,hawking1975}, Bardeen, Carter, Hawking, and Bekenstein demonstrated that asymptotically flat, stationary black hole configurations satisfy
   \be\label{firstlaw}
   dM=T_H\,dS+\Phi\,dQ+\Omega\,dJ\ ,
   \ee
   where $M$ is the black hole's mass, $T_H$ is the temperature of its Hawking radiation, $S$ is its entropy, $Q$ is its total charge, and $J$ is its angular momentum. In the intervening years, it has also been shown that consistently accounting for variations in the cosmological constant $\Lambda$ yields an additional work term $VdP$ in the first law, where $\Lambda$ enters as a thermodynamic pressure with the identification $\Lambda=-8\pi P$ \cite{kastor2009,urano2009}. In the modern viewpoint, the relation \eqref{firstlaw} can be seen as a Gauss law arising from a variation of the Hamiltonian $H_{\xi}$ associated to the phase space flow of the horizon-generating Killing field $\xi^a$, which makes explicit the role of entropy as the Noether charge associated with diffeomorphisms \cite{sudarsky1992,iyer1994}.  Black hole thermodynamics aims (in part) to discern the extent of the similarity between \eqref{1st} and \eqref{firstlaw}, and to uncover the unique insights that emerge where they differ.

   \section*{When is equilibrium plausible?}

   The first law of thermodynamics is a statement of conservation of energy, a concept that becomes immediately subtle in general relativity where defining the mass or energy of spacetime is difficult. The Hamiltonian $H$ for general relativity (indeed any diffeomorphism-invariant theory) vanishes on-shell because it is a pure constraint, so the energy is naively zero for any configuration. However, the variation of $H$ which leads to the Einstein equations necessarily involves the introduction of surface terms which in some cases can be used to define the energy. These definitions of energy typically require additional assumptions about the spacetime, such as asymptotic flatness in the case of the ADM mass, or stationarity in the case of the Komar mass \cite{arnowitt1959,komar1963}. As a result, the interpretation of the quantity $M$ entering into \eqref{firstlaw} varies from one application to the next, and the assumptions entering its construction must be reflected in the circumstance for which the associated first law is being applied.
   \\
   
   Since \eqref{1st} is fundamentally a statement relating equilibrium states it is convenient that in the case of gravity, the mass/energy can be defined unambiguously for stationary spacetimes, which serve as the natural analogue of ordinary equilibrium states. These states play a fundamental role on both sides of the duality, where in the classical theory they also serve as a basis for the {\it H}-theorem, providing a kinetic-theory definition of the fine-grained entropy and the associated second law of thermodynamics \cite{boltzmann1872,lifshits1981}.
   \\
   
   Stationarity is not required for the definition of the ADM mass, which only requires an asymptotically flat spacelike hypersurface $\Sigma$, but is needed for defining the Komar mass, which is essentially just a Killing flux integral at finite radius. Despite this, generalizations of temperature and mass to dynamical spacetimes do exist, along with suggestions for dynamical versions of the first law \cite{hayward1998a}. In such constructions, however, issues are usually encountered when defining an appropriate notion of dynamical entropy. In the covariant framework \cite{iyer1994}, the Wald entropy $S$ appearing in \eqref{firstlaw} is defined through an integral over a cross-section of a Killing horizon, which requires the existence of a bifurcation surface and does not straightforwardly generalize to non-stationary black holes involved in dynamical processes. Furthermore, while the area of the event horizon is invariant under the choice of cross-section in the stationary case, it is no longer so when the horizon is dynamical. Finally, arguing in a dynamical setting that a black hole is `slowly' evolving is delicate due to time-reparametrization invariance of general relativity, though in some cases a geometrically meaningful definition of slowness can be established \cite{booth2004a}. Notions of staticity, equilibrium, and asymptotic structure are thus intimately tied together in the first law, and have the potential to become strained in cosmological and/or dynamical settings, as we will discuss.
   \\
   
   Especially in the context of AdS/CFT and black hole phase transitions, the first law \eqref{firstlaw} is interpreted as an exact differential relating nearby classical solutions with mass $M$, rather than a physical process involving the interaction of a black hole with external matter. This should be obvious from the fact that the ADM/Komar mass cannot change, so different values of $M$ label distinct spacetimes. Nonetheless, it can be appropriate in some circumstances to give a `physical process' interpretation to \eqref{firstlaw} in contrast with this equilibrium interpretation \cite{wald1994a}, though this requires some strict assumptions on the vanishing of terms quadratic in the shear and expansion of the horizon generators. Even still, both interpretations necessarily assume equilibrium initial and final states.
   \\
   
   In the case of black holes, an approximate notion of equilibrium involves arguing that the timescale over which back-reaction from the evaporation process becomes relevant is much longer than the timescale required to assign values to thermodynamic quantities describing the black hole. Establishing whether this is the case requires examining the fundamental nature of Hawking radiation and the measurement process. Hawking radiation not only possesses a quantum-mechanical origin, but also manifests in a distinctly quantum-mechanical way. This can readily be observed for Schwarzschild black holes in asymptotically flat space, which emit Hawking radiation as a greybody at a temperature \cite{hawking1975}
   \be\label{temp1}
   T_H=\frac{\hbar\:\! c^3}{8 \pi k_B G M}\ .\nonumber
   \ee
   The luminosity of the black hole can be approximated using the Stephan-Boltzmann law or by explicit calculation of particle emission rates \cite{page1976a} as
   \be\label{lum}
   L= \sigma A\, T_{H}^4=\frac{\hbar\:\! c^6}{15360\:\! \pi\:\! G^2 M^2} =c^2\,\dfrac{dM}{dt}\ ,
   \ee
   and the peak emission wavelength can be estimated by Wien's displacement law 
   \be\label{lambda}
   \lambda_{\text{max}}=\frac{2\pi \hbar\:\! c}{k_B T_H\left[\mathrm{W}\left(-4 e^{-4}\right)+4\right]}\approx\dfrac{4\pi^2 G M}{c^2}\approx 20 R_g\ ,\nonumber
   \ee
   where $R_g$ is the Schwarzschild radius, $W$ is the Lambert-W function, and we have approximated $\mathrm{W}\left(-4 e^{-4}\right)\approx 0$. The energy and period of the corresponding excitations are simply
   \be
   E_{\text{max}}=\dfrac{2\pi \hbar \:\!c}{\lambda_{\text{max}}}\ ,\qquad T=\dfrac{\lambda_{\text{max}}}{c}\ .\nonumber
   \ee 
   However, the average time between the emission of individual Hawking quanta would be
   \be\label{emm}
   t_{e}\sim \frac{E_{\text{max}}}{L}=\frac{7680\:\! G M}{c^3}\ ,\nonumber
   \ee
   which for a solar mass black hole with $M=M_\odot=1.989\times10^{30} \text{ kg}$ gives
   \be
   t_{e}\approx 0.0206\text{ s}\ ,\qquad\text{while } \quad T\approx 0.0002\text{ s}\ ,\nonumber
   \ee
   and so the mean emission time between individual Hawking quanta is greater than the period of the quanta themselves, a statement that is invariant in the sense that
   \be
   \frac{t_{e}}{T} =\text{const.}\nonumber
   \ee
   It is therefore misleading to imagine---for any size of black hole---a steady stream of classical radiation being observed from the object, except perhaps over exceptionally long timescales. The question of whether quasi-static equilibrium provides an appropriate description then amounts to this: can the temperature $T_H$ be meaningfully assigned to the black hole (in an operational way) before its mass appreciably changes due to evaporation, and is a continuum description valid over the same timescale?
   \newpage
   
   Assume that the observation of $N\gg 1$ quanta suffices to establish a continuum or statistical description of the radiation and determine the temperature to be $T_H$. Assume also some idealized detection apparatus and an asymptotic region in which the particle concept is well-defined. The observation time then required is simply $N*t_e$, which is linear in the black hole mass $M$. This timescale competes with the evaporation timescale, which can be estimated by integrating \eqref{lum} to obtain
   \be
   t_{\text{evap}}= \dfrac{5120\:\!\pi\:\! G^2 M_0^3}{\hbar \:\!c^4}\ ,\nonumber
   \ee
   which is cubic in the initial mass $M_0$ of the black hole. Assume that $N=10^9$, which is comparable to the number of molecules contained in an ideal gas on Earth within a volume of size $10^{-11}\text{ cm}^3$. By the time this number of Hawking quanta are emitted, the {\it fractional} change in the mass of the black hole is 
   \be\label{frac}
   \dfrac{\Delta M}{M_0}\approx \left(\dfrac{dM}{dt}\right)\dfrac{N\:\!t_e}{M_0}=\dfrac{4\times10^{-8}\text{ kg}^2}{M_0^2}
   \ee
   Remarkably this change only becomes appreciable when $M_0\sim 10^{-3}\text{ kg}$, which corresponds to a black hole with an event horizon radius of size
   \be
   R_g= 1.5\times10^{-30}\text{ m}=9.2\times10^5\ l_p\nonumber\ ,
   \ee
   where $l_p$ is the Planck length. Therefore only when the black hole is a few orders of magnitude away from the Planck scale in size does the time required to observe a statistically large number of Hawking quanta become comparable to the timescale over which its own mass appreciably changes, where anyway the approximations entering into Page's law \eqref{lum} and the semi-classical approximation are not expected to be valid. For a solar mass black hole, \eqref{frac} gives a fractional change in mass  of $1/10^{68}$ over the relevant observational time. While the approximations made here are of course crude (the mass-loss is at least 1.8 times greater due to the effective radiating area being larger than just $4\pi R_g^2$, some idealized detection method has been assumed, integration time for long-wavelength modes is ignored, etc.) the quasi-static approximation is nevertheless extremely good for Schwarzschild black holes in asymptotically flat space. While it is plausible that the same is true for more general black hole solutions, adhering to situations where equilibrium is manifest regardless of the timescales involved removes any uncertainty about the effects of back-reaction or dynamics, and is almost surely necessary for holographic applications like AdS/CFT.   
\vspace{-0.5cm}   
   \section*{Problems in de Sitter}
   
   The notion of equilibrium is stressed because it one of the main obstacles arising when moving to a cosmological setting. Today there is incontrovertible evidence for the existence of putative black hole-like objects \cite{abbott2021} and for the accelerated expansion of the Universe \cite{garnavich1998,goldhaber2009a,planckcollaboration2016}, the former implying the existence of Kerr-like black holes, and the latter implying an asymptotically de Sitter geometry as their background. These discoveries urge us to extend our understanding of black hole thermodynamics to asymptotically de Sitter spacetimes, not only for their observational relevance, but also for applications in the emerging field of de Sitter space holography \cite{strominger2001b,larsen2002,balasubramanian2003}.
   \\
   
   Though the first law \eqref{firstlaw} can be readily generalized to cases where $\Lambda\neq 0$, asymptotically AdS black holes (with $\Lambda<0$) seem to lie at the forefront. The relative lack of attention given to de Sitter ($\Lambda> 0$) black hole thermodynamics can be traced to a host of unique issues which are absent in AdS. Perhaps the most salient is the existence of a cosmological horizon, which is known to radiate thermally in analogy with the event horizon \cite{gibbons1977b}. In stationary spacetimes, the temperature of Hawking radiation can be related to the surface gravity $\kappa$ of the generating horizon (in natural units) through
   \be\label{t}
   T_H=\dfrac{\kappa}{2\pi}\ .
   \ee
   Evaluating \eqref{t} for both the cosmological and event horizon, it is easy to show that the cosmological horizon radiates at a much smaller temperature than any black hole that fits within it, necessarily placing the system out of equilibrium due to the heat flux between the two horizons\footnote{For reference, a black hole in thermal equilibrium with the CMB would have $M=4.5\,\text{g}$ and $R_g=67\,\text{$\mu$m}$.}. This is in sharp contrast to asymptotically AdS spaces, which have an attractive gravitational potential and a timelike boundary at infinity that radiation can reach in finite time. These boundary conditions make AdS act as a `box' that confines massless particles, allowing large black holes to come into thermodynamic equilibrium with their own Hawking radiation. 
   \\
   
   \begin{figure}[h]
   	\begin{center}
   		\includegraphics[width=0.4\textwidth]{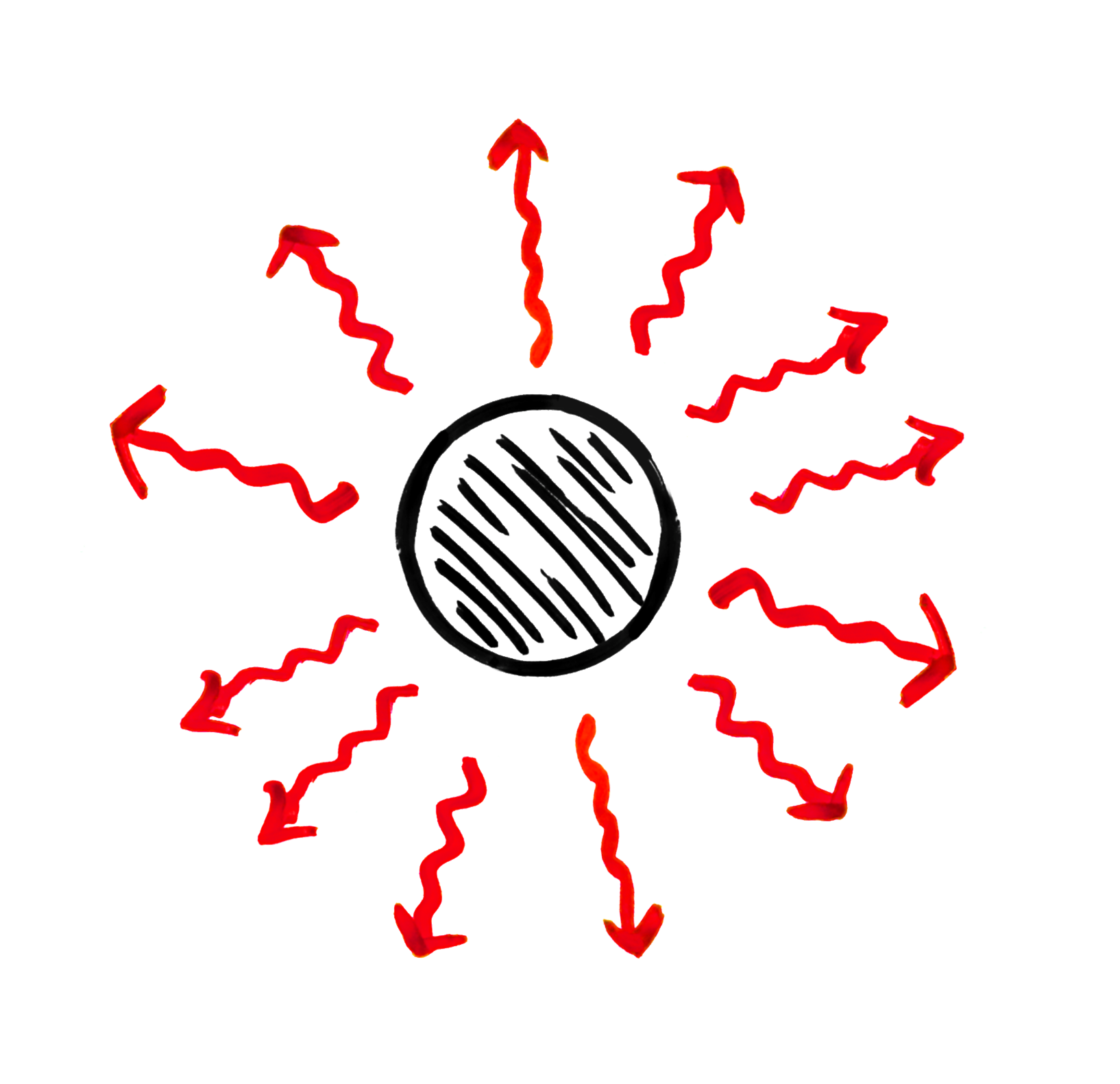}\qquad	\includegraphics[width=0.4\textwidth]{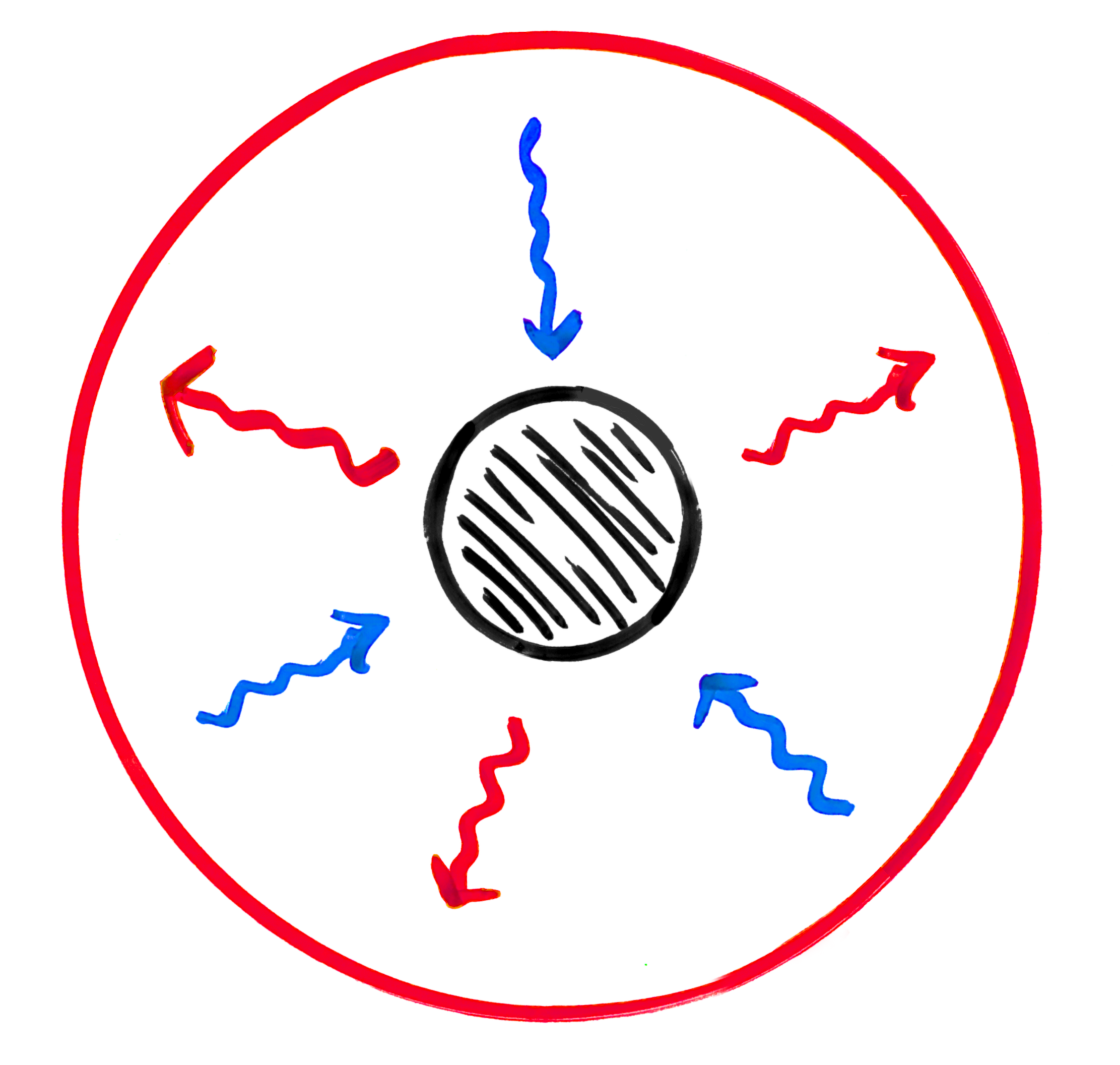}
   		\caption{Evaporating black hole configurations. {\bf Left:} A black hole in asymptotically flat or de Sitter space will eventually evaporate completely. {\bf Right:} In asymptotically anti-de Sitter space or in a cavity, a suitably sized black hole can equilibrate with its own radiation.}
   		\label{fig:evap}
   	\end{center}
   \end{figure}
   
   Another issue with de Sitter spaces is a lack of globally timelike Killing vector field with which to associate the mass, making the construction of conserved charges difficult \cite{bousso2002,balasubramanian2002,anninos2011}. The masses which enter into the usual forms of the first law are defined epacetimes which are asymptotically flat (ADM) or stationary (Komar). In de Sitter, one can recover stationarity by working in the static patch, but then the Killing vector $\xi^a$ that would be used to define the mass becomes spacelike outside the cosmological horizon, rendering the mass conserved in space rather than time \cite{ghezelbash2002c}. In this case $M$ cannot support its usual Noether charge interpretation as being the conserved quantity associated with time-translation invariance. Therefore while a variation resembling the first law exists, the variables entering it do not clearly correspond to the usual thermodynamical ones. The notion of a vacuum state is also problematic in de Sitter, since the global spacetime is non-stationary  \cite{mottola1985,allen1985,goheer2003} and the vacuum state is not even known to be stable \cite{brown1988,anderson2014,firouzjaee2015}.

   \section*{The way forward...}
   
   Various approaches have been developed to circumvent these difficulties, each with their own advantages and drawbacks. One is the {\it effective temperature} approach, where a single temperature $T_{\text{eff}}$ (which depends on both the cosmological and event horizon) is assigned to the entire spacetime \cite{zhangli-chun2010}. This enables one to establish a `first law' which accounts for the presence of both horizons, but suffers from the fact that $T_{\text{eff}}$ lacks a clear physical interpretation and the system still appears out of equilibrium to a local observer. Another approach is to consider only subsets of the parameter space where the two horizon temperatures are equal. Such an approach was adopted in \cite{mbarek2019} for example, where Schwarzschild-de Sitter black holes with conformal scalar hair were examined. In this case equilibrium is trivially established, at the price of being limited to a measure-zero subset of possible configurations. One also requires sufficiently many `charges' to make the temperatures equal, which is not possible for ordinary Schwarzschild-de Sitter black holes.
   \\
   
   One of the more promising directions involves a Euclidean path integral approach, developed first by Gibbons and Hawking \cite{gibbons1977} and elaborated on by York \cite{york1986,braden1990}. This method has a basis in the fundamental relationship between the partition function $\mathcal{Z}$ and the path integral. The partition function for a continuous quantum system with Hamiltonian $H$ at finite temperature $T=\beta^{-1}$  is just
   \be
   \mathcal{Z}=\text{Tr }e^{-\beta H}=\int dq \bra{q}e^{-\beta H}\ket{q}\ ,\nonumber
   \ee
   and from it, thermodynamic quantities like $E$ and $S$ can be easily derived. In the case of black holes, the quantity $\mathcal{Z}$ can be computed from a path integral if one sums over (Euclidean) metrics $g$ which have periodicity $\beta$ in imaginary time $\tau=-it$. Though the measure $\mathcal{D}[g]$ would only be defined formally, in the semi-classical approximation the leading contribution can be computed as
   \be
   \mathcal{Z}=\int_{g(0)}^{g(\tau)}\!\! \mathcal{D}[g]\ e^{-I_E[g]}\approx\sum_{g_{cl}} e^{-I_E\left[g_{c l}\right]}\ ,\nonumber
   \ee
   where $I_E[g]$ is the Euclidean action for the metric $g$, and $I_E\left[g_{c l}\right]$ is the saddle-point contribution from the classical metrics $g_{cl}$ which solve the equations of motion. Although one is summing over Euclidean geometries, the resulting partition function is in fact that of finite-temperature fields in the corresponding {\it Lorentzian} geometry. This Euclidean--Lorentzian relationship is encoded in the Kubo-Martin-Schwinger (KMS) condition \cite{kubo1957,martin1959}, with the Osterwalder-Schrader reconstruction theorem providing the conditions under which such an isomorphism can be established \cite{osterwalder1973}. 
   \\
   
   This method of computing thermodynamic quantities is particularly advantageous for black holes in de Sitter space. One can fix boundary-value data (e.g. the temperature) on a 2-surface at some finite radius $r_c$ between the black hole and cosmological horizons, and compute $\mathcal{Z}$ with the Euclidean metric being cut off at a finite boundary. The specific temperature required so that the Euclidean metric contributions to the path integral are regular depends on both the black hole's charges $\{M,Q,J\}$ and the cosmological constant $\Lambda$, and readily defines an ensemble where the equilibrium temperature is precisely known (it is simply the Hawking temperature blueshifted to the boundary $r_c$). This would physically correspond to placing the black hole in an isothermal cavity, a method which has been applied to a wide variety of black hole spacetimes where a thermodynamic expanse comparably rich to that of AdS has been unveiled \cite{carlip2003,simovic2019,simovic2019a,haroon2020,simovic2021}. 
   \\
   
   Of course, this is far from being the complete story. The stability of the de Sitter vacuum, and the role of de Sitter in holography remain tantalizing open questions. The utility that black hole phase transitions have in uncovering features of strongly coupled systems is continually being uncovered. And where equilibrium thermodynamics {\it can} be formulated, there is a need to examine what kinds of physical situations (if any) correspond to the thermodynamic ensembles involved. Undoubtedly these are important questions, since de Sitter space is an excellent approximation to both the geometry of the current Universe and the early inflationary epoch. Traces of black hole thermodynamics can be found across the landscape of theoretical physics today, and it is exciting to imagine where, in its many guises, it will reveal itself to us next.

 \section*{Acknowledgments}
   
   F.S. is supported by the ARC Discovery Grant No. DP210101279.

\newpage{\pagestyle{empty}\cleardoublepage}

\end{document}